\documentstyle[12pt,axodraw]{article}
\def\pri{^{\, \prime}}

\def\ra{\rightarrow}

\def\prd#1{{\em Phys.~Rev.}~{\bf D#1}\ }

\def\prl#1{{\em Phys.~Rev.~Lett.}~{\bf #1}\ }
\def\plett#1{{\em Phys.~Lett.}~{\bf #1B}\ }
\def\np#1{{\em Nucl.~Phys.}~{\bf B#1}\ }
\def\deg{\ifmmode{^{\circ}}\else ${^{\circ}}$\fi}

\def\gsim{\,\raisebox{-0.13cm}{$\stackrel{\textstyle>}{\textstyle\sim}$}\,}
\def\lsim{\,\raisebox{-0.13cm}{$\stackrel{\textstyle<}{\textstyle\sim}$}\,}
\def\bi{\begin{itemize}}
\def\ei{\end{itemize}}
\def\ed{\end{document}}
\def\be{\begin{equation}}
\def\ee{\end{equation}}
\def\bea{\begin{eqnarray}}
\def\eea{\end{eqnarray}}
\def\beas{\begin{eqnarray*}}
\def\eeas{\end{eqnarray*}}
\def\ns{\nonumber\\[.1in]}
\def\req#1{(\ref{eq:#1})}
\def\eq#1{Eq.~(\ref{eq:#1})}
\def\labeq#1{\label{eq:#1}}

\def\tfrac#1#2{{\textstyle\frac{#1}{#2}}}

\def\thalf{\tfrac{1}{2}}

\def\tthird{\tfrac{1}{3}}

\def\tev{\ \mbox{TeV}}

\def\gsim{\raisebox{-0.5ex}{$\stackrel{>}{\sim}$}}
\def\lsim{\raisebox{-0.5ex}{$\stackrel{<}{\sim}$}}

\def\eb{\end{thebibliography}}

\def\labeq#1{\label{eq:#1}}
\def\req#1{(\ref{eq:#1})}
\def\eq#1{Eq.~(\ref{eq:#1})}
\def\tr{\ifmmode{\mbox{Tr}}\else Tr\fi}
\def\bb{\bibitem}

\def\ms{\ifmmode{M_s} \else $M_s$\fi}
\def\mpr{\ifmmode{{\bar M}_P} \else ${\bar M}_P$\fi}
\def\mpl{\ifmmode{M_P} \else $M_P$\fi}
\def\md{\ifmmode{M_D} \else $M_D$\fi}

\def\al{\ifmmode a_{\ell}\else $a_{\ell}$\fi}
\def\alq{\al^{\rm QED}}
\def\alg{\al^{\rm QG}}
\def\ml{m_{\ell}}
\def\ds{\Delta\sigma}
\def\dsg{\Delta\sigma(\gamma \ell\ra G\ell)}
\def\dssg{\Delta\sigma_{\gamma\ell\ra G \ell}}
\def\order#1{\ifmmode {{\cal O}(#1)}\else ${\cal O}(#1)$\fi}

\def\omd{\order{\ml^2/\md^2}}
\def\tr{R_{6-n}}
\begin{document}
\begin{titlepage}
\begin{flushright}  {\sl NUB-3196/99-Th}\\
hep-ph/9904318
\end{flushright}
\vskip 0.5in

\begin{center}
{\Large\bf Gravitons and the Drell-Hearn-Gerasimov Sum Rule}\\ [.5in] {Haim
Goldberg}\\ [.1in]
{\it Department of Physics\\ Northeastern University\\ Boston, MA 02115, USA}\\
\end{center}
\vskip 0.4in

\begin{abstract}
One-loop diagrams containing a graviton provide a finite contribution to the
anomalous magnetic moment $\al$ of a lepton, whether or not the graviton
propagates in $n$ large extra compact dimensions. In the present work, the tree
graph photoproduction of a graviton, integrated up to an arbitrary cutoff, is
shown to violate the Drell-Hearn-Gerasimov sum rule for $\al^2$. The possibility
of resurrecting the sum rule from high energy contributions originating in
string excitations
is discussed in a qualitative manner, and various problems associated with
such a  program are pointed out.
\end{abstract}
\end{titlepage}
\setcounter{page}{2}

\section{Introduction}
Years ago, Berends and Gastmans \cite{berends} calculated the
one-loop contribution of  virtual massless gravitons to $\al=\thalf
(g-2)_{\ell},$ the anomalous magnetic moment of a lepton $\ell,$ and
obtained a finite result. Recently, Graesser
\cite{graesser} has redone this calculation in the context of a revised picture
of gravity
\cite{add} which is of much current interest: in this version,  the
gravitational sector lives in an expanded
$D=(4+n)$-dimensional spacetime, with the extra $n$ dimensions compactified on
surfaces whose characteristic sizes may be as large as a millimeter. Graesser
has considered the contribution to \al\ of the resulting Kaluza-Klein tower of
spin-2 gravitons (and spin-0 partners). The contribution of each KK mode is again
finite, as is the total if the sum on modes is appropriately cut off at or
near the
$D$-dimensional fundamental scale $\md$ \cite{garousi}. The finiteness of these results raises
an interesting question: is a well-known sum rule for \al$^2,$ the
Drell-Hearn-Gerasimov (DHG) sum rule \cite{dhg}  satisfied in quantum gravity,
at least at the one-loop level? In accord with a general argument by Brodsky
and Schmidt \cite{brodsky},  it will be seen that, at
the one-loop level, this sum rule  requires the
vanishing of  a certain  integral involving tree-level contributions to a
difference of polarized  cross sections for the photoproduction of gravitons. As
a result of a tedious but straightforward computation, it will be shown
that this sum rule is not satisfied,  neither in 4-dimensional nor in
$(4+n)$-dimensional gravity. At one level, this may be a plausible result:
gravity, being a non-renormalizable theory, does not satisfy the finiteness
criteria (to be detailed below) necessary for validity of the sum rule.
{}From a different perspective, the finiteness of the gravitational contribution
to \al\ at the one-loop level, combined with a string-based belief in
Reggeization of amplitudes (including gravitational) at very high energies,
suggest that perhaps the sum rule would regain validity upon inclusion of all
the high energy string excitations. In conformance with the
original $s$- and $t$-channel string duality, this suggested an examination of
the constraints on the Regge behavior in order that high energy contributions
at at least be consistent with restoring the validity of  the sum rule. This
could only be done in a rough and speculative fashion, and a discussion of the
problems associated with imposing the resulting constraints will be discussed in
the context of Type I$\pri$ string theories. Speculations aside, the principal
concrete result remains that, to lowest order, the DHG sum rule is not obeyed by
the tree-level contributions involving photoproduction of gravitons.

\section{The DHG Sum Rule and Quantum Gravity}
Under certain conditions on the high energy behavior of both the real and
imaginary part of the forward spin-difference Compton amplitudes, there exists
a sum rule for $\al^2.$ For a spin-1/2 target, it reads \cite{dhg}

\be
\al^2\ =\ \frac{\ml^2}{2\pi^2\alpha}\ \int_{\nu_{th}}^{\infty}\
\frac{d\nu}{\nu}\
\Delta\sigma(\nu)\ \ ,
\labeq{dhg}
\ee
where $\Delta\sigma(\nu)\equiv \sigma_P(\nu)-\sigma_A(\nu)$ is the difference
between photoabsorption cross sections for the scattering of a photon of lab
energy $\nu$ from a target lepton in the cases where the initial photon and
lepton spin components along the incident photon direction are parallel and
anti-parallel, respectively. The validity of the sum rule is predicated on both
the vanishing of $\ds$ at high energies and on the absence of polynomial terms
in the real part of the forward Compton spin-difference amplitude $f_2$. Both
conditions would obtain if the full Compton amplitude were to Reggeize
\cite{jaffe}; the details of this Reggeization at string energies will be
critical in resolving the problem raised by the calculation which follows.

In the Standard Model, the validity of the sum rule requires the vanishing of
the integral on the RHS of \eq{dhg} when the cross section is calculated from
the lowest order tree graphs, namely $\gamma+\ell\ra\gamma+\ell,\;
W^-+\nu_{\ell}$ \cite{altarelli}. This follows from the observation that the
LHS of \req{dhg} when calculated in the Standard Model is of \order{\alpha^2}
(for fixed $\sin\theta_W),$ whereas in Born approximation the RHS is of
\order{\alpha}, and hence must vanish. This is indeed found to be the case, as
shown in the explicit calculation of Altarelli, Cabibbo, and Maiani
\cite{altarelli}. Using loop expansion techniques, the result has been
generalized by Brodsky and Schmidt \cite{brodsky}: the validity of the DHG sum
rule requires the vanishing of the integral on the RHS for the sum of $2\ra 2$
processes $\gamma\ell\ra bc$ in the Born approximation. This then raises the
interesting question: since the one-loop quantum gravity contributions to $\al$
are finite and calculable, is the DHG sum rule satisfied to the appropriate
order in this case? If not, can we learn anything from the failure to satisfy?

In order to simplify the discussion, I will limit the physics to QED and
gravity (the rest of the Standard Model can be included with little
complication). In this case, at the one loop level, the corrections to $\al$
{}from these sources are additive:
\be
\al=\alq+\alg\qquad \mbox{\em(1 loop)}\ \ .
\labeq{oneloop}
\ee
Thus, to one-loop level, the sum rule may be written as

\be
\left({\alq}\right)^{\ 2}+2\ \alq\ \alg\ + \left({\alg}\right)^{\ 2}\
= \frac{\ml^2}{2\pi^2\alpha}\
\int_{\nu_{th}}^{\infty}\ \frac{d\nu}{\nu}\
\left(\left.\Delta\sigma(\nu)\right|_{\rm QED}+\Delta\sigma\pri(\nu)\right)\ \ .
\labeq{add}
\ee
The first term under the integral is the pure QED contribution, up to the
appropriate order, and the second is the mixed gravity-QED cross section. Since
QED by itself obeys the DHG sum rule, the first terms on the LHS and RHS will
cancel.
With this
cancellation, and to lowest order in $\ml^2,$ the sum rule reads
\be
2\ \alq\ \alg\ + \left({\alg}\right)^{\ 2}\
= \frac{\ml^2}{2\pi^2\alpha}\
\int_{s_{th}}^{\infty}\ \frac{ds}{s}\
\Delta\sigma\pri(s)\ \ .
\labeq{adda}
\ee
I will now consider separately the cases with gravity propagating in $n>0
\ (D>4)$ and $n=0\ (D=4)$ dimensions, respectively.\bigskip

\noindent\underline{\boldmath {$n>0$}}\bigskip

\noindent I first briefly review the properties of $4+n$-dimensional gravity I
will need for the discussion which follows. For compactification on an
$n$-torus, the
$D$-dimensional Planck scale
$\md$ is related to the large radius $R_n$ and the reduced Planck mass via
\cite{add,addone,giud}
\be
(\mpr/\md)^2 = (\md R_n)^{n}\ \ ,
\labeq{msr}
\ee
so that $\md$ can range from $\sim \tev$ to $M_P$ for $R\le$1 mm as long as
$n\ge 2.$ The resulting Kaluza-Klein tower of spin-2 gravitons (and spin-0
partners, the radions) can have very small mass splitting $\Delta
m^2=(1/R_n)^2.$ This scale also characterizes the surface tension of the soliton
(say a D-brane) to which is tied the open string containing ordinary matter.
Feynman rules have been developed \cite{giud,lykk} for these couplings, and a
large number of authors have explored the phenomenological implications of this
view of gravity.

Thus, on  the RHS of \req{adda}, the lowest order contributions are the tree
processes
$\gamma\ell\ra G \ell,$ $\gamma\ell\ra \Phi\ell$ where $G$ is one of the spin-2
graviton, and $\Phi$ is its scalar partner (the radion). The cross section for
the latter process is suppressed by a factor of $\ml^2,$ so it does not enter
the present consideration.\footnote{In accordance with the scaling argument in
\cite{brodsky}, the radion contribution should cancel when combined with
\order{\ml^2}\ corrections to the tree-level $\dsg.$}

The calculation of $\dsg$ is straightforward \cite{peskin}. I fix the
kinematics so that the photon is incident along the $+z$-axis with momentum
$q,$ helicity +1, the lepton with momentum $p_1$ along the $-z$-axis. For
massless leptons, the amplitude
\req{ampl} is helicity conserving, and the polarization amplitudes for
$\gamma\ell\ra G\ell$ are then written as
\be
{\cal M}_{P(A)}=\bar u_{L(R)}(p_2)\ {\cal O}^{\mu\nu\rho}\ u_{L(R)}(p_1)
\ \epsilon_{\rho}^{(+1)}(q)\ \left(\epsilon_{\mu\nu}^{\Lambda}(k)\right)^*\ \ ,
\labeq{em}
\ee
Squaring and summing over the final state graviton helicity, one finds (for all
particles massless except for the graviton, with mass $m$) the differential
cross sections in the $c.m.$
\be
d\sigma_{P(A)}/d\cos\theta=\displaystyle\sum_{\Lambda}\left|{\cal
M}_{P(A)}\right|^2/(8\pi\sqrt{s})^2\ \cdot (1-x)\ .
\labeq{diffa}
\ee
where $x\equiv m^2/s,$ and
\be
\displaystyle\sum_{\Lambda}\left|{\cal
M}_{P(A)}\right|^2 =
(\epsilon_{\rho\prime}^{(+1)}(q))^*\mbox{Tr}\left(\widetilde{\cal
O}^{\mu\pri\nu\pri\rho\pri}p_2\hspace{-10pt}/\ \ {\cal
O}^{\mu\nu\rho}p_1\hspace{-10pt}/\ \ (1+(-)\gamma_5)/2\right)\
\epsilon_{\rho}^{(+1)}(q)\ {\cal P}_{\mu\pri\nu\pri;\mu\nu}(k)\ \ ,
\labeq{diffb}
\ee
The graviton spin-2 projection operator ${\cal P}$ is given in
Refs.\cite{giud,lykk}, and $\widetilde {\cal O}=\gamma^{0\dagger}{\cal
O}^{\dagger}\gamma^0.$ Finally, from
\req{diffa} and \req{diffb},
\bea
\frac{d\sigma_P}{d\cos\theta}-\frac{d\sigma_A}{d\cos\theta}
&=&(\epsilon_{\rho\prime}^{(+1)}(q))^*
\ \mbox{Tr}\left(\widetilde{\cal
O}^{\mu\pri\nu\pri\rho\pri}p_2\hspace{-10pt}/\ \ {\cal
O}^{\mu\nu\rho}p_1\hspace{-10pt}/\ \ \gamma_5\right)\
\epsilon_{\rho}^{(+1)}(q) \ \cdot\ns
&& \quad{\cal P}_{\mu\pri\nu\pri;\mu\nu}(k)\ \cdot (1-x)\ /\ (8\pi\sqrt{s})^2
\ \ .
\labeq{pol}
\eea

The contributing diagrams are given in Refs.\cite{giud,lykk} and are shown in
Fig.1.
\bigskip

\begin{center}
\SetScale{.7}
\begin{picture}(540,160)(0,0)
\SetOffset(80,0)
\ArrowLine(0,0)(90,0)
\Vertex(90,0){1.5}
\ArrowLine(90,0)(180,0)
\Photon(0,60)(90,60){1.5}{5}
\Vertex(90,60){1.5}
\Photon(90,0)(90,60){1.5}{5}
\Photon(90,60)(180,60){2}{5}
\Photon(90,60)(180,60){-2}{5}
\SetOffset(250,0)
\ArrowLine(0,0)(90,0)
\Vertex(90,0){1.5}
\ArrowLine(90,0)(180,0)
\Photon(0,60)(90,0){1.5}{8}
\Photon(90,0)(180,60){2}{8}
\Photon(90,0)(180,60){-2}{8}
\SetOffset(80,110)
\ArrowLine(0,0)(60,0)\Text(0,-5)[lt]{$p_1$}
\Vertex(60,0){1.5}
\Line(60,0)(120,0)
\Vertex(120,0){1.5}
\ArrowLine(120,0)(180,0)\Text(130,-5)[rt]{$p_2$}
\Photon(0,60)(60,0){1.5}{6}\Text(0,45)[lb]{ $q$}
\Photon(120,0)(180,60){2}{6}
\Photon(120,0)(180,60){-2}{6}
\Text(125,45)[lb]{$k$}
\SetOffset(250,110)
\ArrowLine(0,0)(60,0)
\Vertex(60,0){1.5}
\Line(60,0)(120,0)
\Vertex(120,0){1.5}
\ArrowLine(120,0)(180,0)
\Photon(0,60)(120,0){1.5}{10}
\Photon(60,0)(180,60){2}{10}
\Photon(60,0)(180,60){-2}{10}
\end{picture}
\end{center}\bigskip

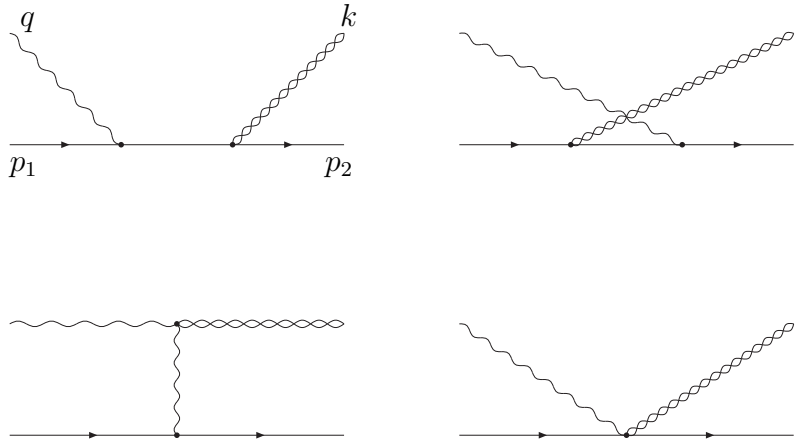
\centerline{ Figure 1: {\em Graphs contributing to}
$\gamma\ell\ra G\ell$.}\bigskip

\noindent The amplitude can be obtained from these references,
and I write it here for completeness (kinematics in Fig.1):
\bea
i{\cal O}_{\mu\nu\rho}&=&
\left(\frac{ieQ_{\ell}}{4\mpr}\right)
\left[
\gamma_{\mu}(P+p_2)_{\nu}
\ (P\hspace{-7pt}/\ /s)\ \gamma_{\rho}+\gamma_{\rho}\ (K\hspace{-7pt}/\ /u)
\ \gamma_{\mu}(p_1+K)_{\nu}\right]\ns
&&- \left(\frac{ieQ_{\ell}}{\mpr}\right)\left[
q_{\mu}Q_{\nu}\gamma_{\rho}+(q\cdot Q)
\eta_{\mu\rho}\gamma_{\nu}
-\eta_{\mu\rho}Q_{\nu}q\hspace{-6pt}/\
-\gamma_{\mu}q_{\nu}Q_{\rho}\right]/t\ns
&&-\left(\frac{ieQ_{\ell}}{2\mpr}\right)\gamma_{\mu}\eta_{\nu\rho}\quad +
\mu\leftrightarrow\nu\ \ .
\labeq{ampl}
\eea
In \eq{ampl}, $Q_{\ell}$ is the charge on the lepton, $\mpr=(8\pi G_N)^{-1/2},\
P=p_1+q,\ Q=k-q,\ \mbox{and}\ K=p_1-k$ (see Fig. 1).

{}From \req{pol} and \req{ampl} I find

\bea
\frac{d\sigma_P}{d\cos\theta}-\frac{d\sigma_A}{d\cos\theta}&=&\frac{1}{16}\ \frac{\alpha}
{\mpr^2}\ (1-x) \ \left[ \left(\frac{s}{t}\right)( 4x^2 - 8x ) +
\left(\frac{s}{u}\right)(x- 2x^2)\right.\ns
&&\left. +\ (2x^2-4x-4 ) + \left(\frac{u}{s}\right)(4-x )\right]\ \ ,
\labeq{dsig}
\eea
where $t=Q^2,\ u=K^2.$ The cross section has a smooth massless limit $(x=0)$
which coincides with that obtained from starting with massless gravitons.

With the substitutions $(t,u)=-\thalf s(1-x)(1\mp \cos\theta),$ and the
imposition of a collinear cutoff
 $-1+\delta\le
\cos\theta\le 1-\delta$, the total polarization cross section is obtained
by integrating \req{dsig} over $\cos\theta:$

\be
\dssg(s,m^2)=\frac{1}{16}\ \frac{\alpha}
{\mpr^2}\ \left[\log\left(\frac{2}{\delta}-1\right)( 14x - 4x^2 ) - (12
-9x-6x^2+3x^3)\right]\ \ .
\labeq{cs}
\ee

The contribution to $\dssg(s)$ from an entire KK tower of gravitons is given
(approximately) by integrating over the density of states\
\cite{giud,lykk,peskin} up to the kinematic limit

\bea
\dssg(s)&=&\frac{2\pi^{n/2}}{\Gamma(n/2)}\ R^n\
\int_0^{\sqrt{s}\ }\dssg\ (s,m^2)\ m^{n-1}\ dm\ns
&=&\frac{\pi^{n/2}}{\Gamma(n/2)}\
\frac{\mpr^2}{\md^2}\left(\frac{s}{\md^2}\right)^{n/2}\ \int_0^1\dssg\ (x)\
x^{n/2-1}\ dx\ns &=&\frac{1}{16}\
\frac{\alpha}{\md^2}\cdot\ \left(\frac{s}{\md^2}\right)^{n/2}\ A_n\ \ ,
\labeq{intm}
\eea
where, {\em e.g.,}
\beas A_2&=&\pi\left(-25/4+(17/3)\log(2/\delta-1)\right)\\
A_4&=&\pi^2\left(-21/10+(11/3)\log(2/\delta-1)\right)\eeas

Finally, the contribution of \req{intm}, integrated up to some upper limit
$\bar s\le \md^2$ to the RHS of the sum rule
\req{adda} is

\be
\frac{\ml^2}{2\pi^2\alpha}\ \int_{s_{th}}^{\bar s}\ \frac{ds}{s}\ \dssg(s)=
\frac{1}{16\pi^2}\  \left(\frac{\ml^2}{\md^2}\right)\
\left(\frac{A_n}{n}\right)\ \kappa^{n/2}\ \ ,
\labeq{dhgb}\ee
where $\kappa\equiv \bar s/\md^2.$

\noindent On the LHS of \req{adda}, $\alq\sim\order{\alpha}$ and
(using the same cutoff at $\bar s)$ $\alg\sim\kappa^{n/2}\omd$ \cite{graesser};
thus
\req{dhgb} by itself violates the DHG sum rule at the one-loop level. Before
discussing this result, I will present the
analogous result for the case with no extra dimensions.

\noindent\underline{\boldmath $n=0$}\bigskip

\noindent As mentioned after \eq{dsig}, the $D=4$ case is obtained by setting $x=0$ in
that equation, or in \eq{cs}:
\be
\dssg(s)=\dssg(s,0)=-\frac{3}{4}\ \frac{\alpha} {\mpr^2}=-\frac{6\pi\alpha}{\mpl^2}\ \ .
\labeq{csfour}
\ee
Then, integration up to $\bar s\le \mpl^2$ gives a contribution to the RHS of
the sum rule
\be
\frac{\ml^2}{2\pi^2\alpha}\ \int_{s_{th}}^{\bar s}\ \frac{ds}{s}\ \dssg(s)=
-\frac{3}{\pi}\ \frac{\ml^2}{\mpl^2}
\cdot\  \log\left(\frac{\bar s}{s_{th}}\right)
\ \ .
\labeq{dhgbfour}
\ee
Again, since $\alq\sim\order{\alpha}$ and $\alg\sim\order{\ml^2/\mpl^2}$
\cite{berends}, this contribution by itself violates the sum rule \req{adda}.

I now turn to discuss these results.
\section{Discussion}
The failure of the gravitational contributions integrated to a large scale
$\bar s$ to satisfy the DHG sum rule at the one-loop level suggests at least two
possibilities: (1) the sum rule is not valid for processes involving quantum
qravity  (2) there are contributions from $s>\bar s$ which cancel the low
energy contribution and render the sum rule valid.

As mentioned in the introduction, and discussed at length in Ref. \cite{jaffe},
the sum rule can fail because the spin-difference forward Compton amplitude
$f_2(s)$ has fixed poles ({\em i.e.,} a polynomial piece to the real part), or
an imaginary part whose asymptotic behavior requires a subtraction for the
dispersion relation. One or both of these is certainly possible: for example,
in the $n=0$ case, there may be at order $\alpha/\mpl^2$ a
gravity-induced ``seagull'' term in
$f_2.$ It is not immediately apparent, however, how this can cancel
against the logarithmic cutoff in \req{dhgbfour}. As far as the imaginary part is
concerned, one may certainly conceive of spin-dependent
gravitational contributions ({\em e.g.,} spinning black holes) whose
contributions vitiate the sum rule. In such cases, within the present state
of knowledge about non-perturbative quantum gravity, there is  not much more
to say. It is then interesting to speculate about the second possibility above,
the possibility of cancellation.\bigskip

\noindent {\large {\bf Possible Role of String Theory}}\medskip

\noindent String theory
suggests that $f_2$  Reggeizes for
$s$ larger than the string scale
$\ms^2,$ at once eliminating the possibility of the fixed poles and the
non-convergence. Moreover, the required Reggeization takes place at the
tree-level (the Veneziano amplitude being equivalent to a sum of poles), so that
according to the loop-counting criteria discussed in \cite{brodsky}, the
additive form of
\req{adda} is valid. I will discuss in turn the two cases of
large extra dimensions $(n>0)$ and no large extra dimensions $(n=0)$.\medskip

\noindent {\boldmath $n>0$:} In this case, I will adopt the string
description presented in \cite{ant}, with reference to previous work in
\cite{lykken} and \cite{polchwitt}, and detailed in \cite{addone,tye}. This is
based on a type I theory of open and closed strings, with a $T$-duality
transformation allowing a large radius in the $n$ extra dimensions, as well as
a weak coupling description. Matter resides in open strings tied to D3-branes
and $\lambda=2\alpha$ serves as the string coupling expansion parameter
\cite{ant}. The 4-point open string Compton amplitude is then written as in
\cite{green}, with the appropriate kinematic factors assuring the behavior (for
$s\gg\ms^2)$
\be
\mbox{Im} f_2(s)\ \sim\ \alpha\ \cdot\ const,\quad \Delta\sigma\pri\sim \alpha/s
\labeq{asym}
\ee
which allows convergence. The contribution to the DHG sum rule from the Regge
region is
\be
\frac{\ml^2}{2\pi^2\alpha}\ \int_{\bar s}^{\infty}\
\frac{ds}{s}\ \Delta\sigma\pri(s)\sim
\left(\frac{\ml^2}{\md^2}\right)\ \left(\frac{1}{\kappa}\right)\ \ .
\labeq{regge}
\ee
A necessary (but certainly not sufficent!) condition that \req{regge} could
cancel the lower energy integral
\req{dhgb} is that
$\kappa\sim
\order{1}$, {\em i.e.,} the perturbative treatment of $\gamma \ell\ra G\ell$ is
reliable for $\sqrt{s}\ \lsim\ \md.$ Some discussion of this point is given in
\cite{giud}. An important ingredient is that the D-brane surface tension, which
controls the ``soft'' scale for gravitons emitted transverse to the brane, is
larger than the string scale \cite{garousi}, so that $\md$ may be reasonable
as an energy cutoff for the perturbative treatment of graviton emission.
The requirement $\bar s\sim \md^2$ {\em vs.} the expected $\bar
s\sim \ms^2$ is perhaps disconcerting, and some discussion will be presented in
a following section. All of  this is a long way from claiming that
cancellation occurs: for
example, the  the collinear cutoff dependence (which occurs only in this
extra-dimension case) in
\req{intm} has no counterpart in \req{regge}. One is assured by the
Lee-Nauenberg theorem \cite{lee} that including all channels will regulate such
singularities to measurable quantities, but how (and whether) this happens
smoothly over the whole energy range in the present case certainly remains to
be shown.\medskip

\noindent {\boldmath $n=0$:} I use the same model here as above, except with no
large dimensions. In addition to $\lambda=2 \alpha,$ one also has (after the
T-duality transformation on all six compact coordinates)\cite{tye}
\be
\ms=(1/\sqrt{2})\alpha\mpl/r^3\ \ ,
\labeq{msmp}
\ee
where $r\ge 1,$ the compactification radius in units of $\ms^{-1},$ can be used
to adjust $\ms=M_{GUT}.$ The Compton amplitude in the Regge region is the same
as in \req{asym}, so that the contribution from $s\ge \bar s$ to the DHG
integral is given by
\be
\frac{\ml^2}{2\pi^2\alpha}\ \int_{\bar s}^{\infty}\
\frac{ds}{s}\ \Delta\sigma\pri(s)
\sim\frac{\lambda}{\alpha}\ \frac{\ml^2}{\bar s}
\sim\left(\frac{\ml^2}{\mpl^2}\right)\left( \frac{\mpl^2}{\bar s}\right)\ \ .
\labeq{reggefour}
\ee
In this case, cancellation of the lower energy contribution \req{dhgbfour} is
possible only for $\bar s\sim\mpl^2\ \gsim\ \ms^2/\alpha^2.$ Again, this is
merely a necessary condition -- there is no demonstration of cancellation, and
problems attached to it as well: for example, the  logarithmic
cutoff factor in \req{dhgbfour} does not appear in the purported cancelling
term \req{reggefour}.\bigskip

\noindent {\large {\boldmath $\md$ (or $\mpl$)  {\em vs.} $\ms$}}\medskip

\noindent To pursue the possibility of cancellation, one must
reconcile the appearance of $\bar s=\md^2 (\mpl^2)$ with the expected $\bar s=
\ms^2$ as the cutoff. As discussed above, there may be plausible arguments why
graviton emission into the bulk may be adequately described by perturbation
theory for energies up to $\md;$ however, it is difficult to understand why the
open string contribution to Compton scattering would be delayed for energies
above $\ms,$ until $\md.$ It might be thought that the difference is not
significant, since in the string theory model $\ms$ and $\md$ are numerically
close \cite{ant,addone}; however, this is not the case. The insertion of $\bar
s=\ms^2$ as the cutoff would cause a mismatch of a factor of $\alpha^2$ between
the contributions below and above $\bar s:$ in the case $n>0,$ this occurs when
the relation $\md\sim\alpha^{-2/(n+2)}\ms$ \cite{ant,addone} (along with $\bar
s=\ms^2)$ is inserted into
\req{dhgb} and $\bar s=\ms^2$ in \req{regge};  for $n=0,$ one has a
similar situation, with $\mpl\sim \ms/\alpha$ in \req{dhgbfour} and $\bar
s=\ms^2$ in
\req{reggefour}. There is only one parameter available to play with, namely
the compactification radius $\tr$ of the $6-n$ remaining dimensions in the
bulk. Can this be of a form such as to remove this parametric disparity between
$\md(\mpl)$ and $\ms?$ Consider again the Type I$\pri$ theory
\cite{ant,addone,tye}, with the standard model fields residing in open strings
tied to a D3-brane, and gravity propagating in the 10-dimensional bulk.
Allowing (as before) $n$ of the compact dimensions to be large, of equivalent
toroidal radius $R_n$, and $6-n$ to be small (of radius $\tr\sim \ms^{-1}$),
one obtains the relation \cite{ant,addone}
\be
(\md/\ms)^{n+2}=\frac{1}{4\pi\alpha^2}\ (\tr\ms)^{6-n}\ \ .
\labeq{trms}
\ee
Thus, for $\alpha\sim 0.1$   it is possible to simultaneously have $\tr\ms>1$
and $\md\sim \order{\alpha^0}\ms$ if
\be
(\tr\ms)^{6-n}\ \gsim\ 100\ \alpha^2\ =\ 25\ \lambda_{I\pri}^2\ \ .
\labeq{six}
\ee
where $\lambda_{I\pri}$ is the Type I$\pri$ string coupling constant. This is a
statement that the compactification scale is parametrically tied to the dilaton
expectation value \cite{dine}. It is not meant to be a perturbative statement,
but as a constraint on the non-perturbative minima in the $S-T$ modular space.
Needless to say, this is very {\em ad hoc}
-- I only present it as a hypothetical way of reconciling the $\ms/\md$
problem.

This scenario is more circumscribed when there are  no  large extra
dimensions. In that case one has, in analogy to
\req{trms}, $(\mpl/\ms)^2=(2/\alpha^2)(R_6\ms)^6$\ \cite{tye}. However, now  $\ms$
is identified with $M_{GUT},$ so that $\mpl/\ms\simeq 500.$ If this be due to
modular geometry rather than a factor of $\alpha,$ we must impose
$(R_6\ms)^6\simeq 30,000\ \lambda_{I\pri}^2$ on the compactification volume, or
$R_6\ms\simeq 6\
\lambda_{I\pri}^{\tthird},$ on the compactification radius.

\section{Concluding Remarks}

The finiteness of the one-loop gravitation contribution to the anomalous
magnetic moment \al\ led naturally to the question of whether the DHG sum rule for
\al$^2$ is
satisfied. The failure to satisfy the sum rule for perturbative contributions
below an arbitrary scale $\bar s$ could be ascribed to the failure to understand
quantum gravity in the strong coupling (high energy) region. Alternatively,
there were presented some speculations concerning
possibly compensating contributions above the string scale.
In both cases $(n>0\ \mbox{and}\ n=0)$ this possibility was beset with
uncertainty  concerning various logarithmic cutoff factors, not to speak of
the absence of an exact calculation. In the simplest (Regge) approximation,
the possibility of a string `fix' for the
validity of the sum rule imposed a {\em necessary} parametric condition:
namely, that the ordinary tree level perturbative contribution be included all
the way to the respective Planck scales $\md,\ \mpl$ (rather than the expected
string scale $\ms),$ and that the string Compton amplitude used thereafter.
Identifying  the string and Planck scales is possible through a certain
dependence of the compactification volume for the $6-n$ `small' dimensions on
the dilaton expectation value. The
simple analysis presented here, using only an open string Compton amplitude,
ignores specifically non-perturbative gravitational contributions (such as
black hole formation \cite{nussinov}) which may set in at $\md\ (\mpl)$. Perhaps
these do not contribute to the forward spin-difference amplitude
$f_2$ (the leading graviton trajectory considered in \cite{nussinov}
does not contribute to $f_2(0)$);
however, no such statement is possible in the presence of spinning black holes.
It is certainly an open question whether these have been eliminated in the
strong-weak duality transformation particular to the model considered.

In sum, the principal finding is that the DHG sum rule is not satisfied at the
one-loop level. Although one may speculate
along certain stringy fixes to this situation, there is at present no
compelling reason to adopt these in preference to simply pleading ignorance
about the convergence properties of $f_2$ in the non-perturbative regime of quantum gravity.

\subsection*{Acknowledgement} I would like to thank Zurab Kakushadze and Tom Taylor
for helpful conversations. This research was supported in part by the National
Science Foundation through Grant No. PHY-9722044.

\begin{thebibliography}{99}
\bb{berends}F.A. Berends and R. Gastmans, \plett{55}(1975) 311.
\bb{graesser}M.L. Graesser, ``Extra Dimensions And The Muon Anomalous Magnetic
Moment'', hep-ph/9902310.
\bb{add}N. Arkani-Hamed, S. Dimopoulos, and G. Dvali, \plett{429}(1998) 263.
\bb{garousi}M.R. Garousi and R.C. Myers, \np{475}(1996) 193;\\
A. Hashimoto and I.R. Klebanov, \plett{381}(1996) 437;
J. Polchinski, S. Chaudhuri, and C.V. Johnson, ``Notes on
D-branes'', NSF-ITP-96-003, hep-th/9602052;\\
R. Sundrum, \prd{59}(1999) 085009.
\bb{dhg}S.D. Drell and A.C. Hearn, \prl{16}(1966) 908;\\ S.B. Gerasimov, {\em
Sov. J. Nucl. Phys.} {\bf 2} (1966) 430. \\
For a recent application, see, {\em
e.g.,} S.J. Brodsky, T.G. Rizzo and I. Schmidt,
\prd{52}(1995) 4929. Subsequent to  the appearance of the present work,
T.G. Rizzo, hep-ph/9904380 has discussed  possible modifications to the DHG
sum rule for $\gamma\gamma\rightarrow W^+W^-$ due to graviton exchange
\bb{brodsky}S.J. Brodsky and  I. Schmidt, \plett{351}(1995) 344.
\bb{jaffe}R.L. Jaffe and Z. Ryzak, \prd{37}(1988) 2015.
\bb{altarelli}G. Altarelli, N.Cabibbo, and L.Maiani, \plett{40}(1972) 415.
\bb{addone}N. Arkani-Hamed, S. Dimopoulos, and G. Dvali, \prd{59}(1999)086004.
\bb{giud}G.F. Giudice, R. Rattazzi, and J.D. Wells, \np{544}(1999) 3.
\bb{lykk}T. Han, J.D. Lykken, and R.~-J. Zhang, \prd{59}(1999)105006.
\bb{peskin}Cross sections for related processes have
been calculated in Refs. \cite{giud,lykk}, and by E.A. Mirabelli,
M. Perelstein and M.E. Peskin, \prl{82}(1999) 2236.
\bb{ant}I. Antoniadis, N. Arkani-Hamed, S. Dimopoulos, and G. Dvali,
\plett{436}(1998) 257.
\bb{lykken}J.D. Lykken, \prd{54}(1996) 3693.
\bb{polchwitt}J. Polchinski and E. Witten, \np{460}(1996) 525.
\bb{tye}Additional helpful discussion can be found in G. Shiu and S.-H.Henry
Tye, \prd{58}(1998) 106007; Z. Kakushadze and S.-H. Henry Tye, \np{548}(1999) 180.
\bb{green}M.B. Green, J.H. Schwarz and E. Witten, {\em Superstring
Theory}, Vol. 1, p. 420, Cambridge University Press (1987).
\bb{lee}T.D. Lee and M. Nauenberg, {\em Phys. Rev.} {\bf 133B} (1964) 1549.
\bb{dine}This type of parametric dependence, where $V_{compact}\sim \lambda^2$
in the heterotic sector, has been discussed in another context by M. Dine and
Y. Shirman, \plett{377}(1996) 36.
\bb{nussinov}S. Nussinov and R. Shrock, \prd{59}(1999) 105002.
\eb\ed
